\documentstyle{article}
\begin{document}

\baselineskip=23pt

\begin{center}
{\bf {\Huge Twin-jets as a potential distance detector}} \footnote{%
This work was subsidized by the Special Funds for Major State Basic Research
Projects and by National Natural Science Foundation of China.}

\vspace{6mm}

{\bf Yi-Ping Qin$^{1,2,3,4}$ and Guang-Zhong Xie$^{1,2,4}$ }

{\bf $^1$Yunnan Observatory, Chinese Academy of Sciences, Kunming, Yunnan
650011, P. R. China}

{\bf $^{2}$ National Astronomical Observatories, Chinese Academy of Sciences 
}

{\bf $^{3}$ Chinese Academy of Science-Peking University joint Beijing
Astrophysical Center }

{\bf $^{4}$ Yunnan Astrophysics Center }

\vspace{6mm}

{\bf {\Large ABSTRACT}}
\end{center}

A preview study of using observational data of proper motions and Doppler
shifted velocities of twin-jets to determine the distance of sources inside
and outside our galaxy is made. We investigate the feasibility of this
method by studying the uncertainty of the distance caused by the
uncertainties of the measured quantities. It shows that, when the motion of
components of the jet is relativistic, then the uncertainty of the distance
is within the same order of the uncertainties of the measured values of
proper motions and Doppler shifted velocities. In particular, when assuming
the pattern speed equals the flow speed in the jet, for $10\%$ uncertainties
of the measured quantities, the uncertainty of the distance caused by them
would be well within $13\%$. With current technique, this method is
realizable. For the convenience of choosing sources to observe, some sources
as potential targets are also listed in this paper.

{\bf Key words:} distance scale --- galaxies: active --- galaxies: jets ---
galaxies: nuclei

\vspace{2mm}

\section{Introduction}

Distance measurement is a long-lived task in astronomical science. The
distance of faraway objects can be determined if its real size or luminosity
is known. This strongly relies on the discovery of some standard candles or
rods. For example, the peak brightness of type Ia supernovae has been served
as a distance indicator and was used to determine the Hubble constant (see,
e.g., Brach 1992). So far, the most reliable measurement of the distance for
extragalactic sources might come from the observation of NGC 4258. Being
observed with VLBI, direct measurement of orbital motions in a disk of gas
surrounding the nucleus of this galaxy was made by tracking individual maser
features from 1994 to 1997, where proper motions of masers as well as
Doppler shifted velocities were obtained (Herrnstein et al. 1999).

VLBI observations have revealed the typical core-jet structure for many
AGNs. According to the unified schemes, an AGN possesses in its compact core
a supermassive black hole surrounded by an accretion disk, and it is always
assumed that there is a twin-jet beamed from the center of the core along
the axis of the disk (see, e.g., Urry and Padovani 1995). For a relativistic
beaming, there will be a Doppler boosting for the flux of components of the
jet, with $f(\nu )=\delta ^pf_0(\nu )$, where $f_0(\nu )$ is the flux
density which would be observed in a reference frame moving with the jet
material, $f(\nu )$ is the flux density observed by us, $\delta =\sqrt{%
1-\beta ^2}/(1-\beta \cos \theta )$, and $p$ can take on the values $%
2+\alpha $ (for a continuous jet) and $3+\alpha $ (for a discrete component)
for $f(\nu )\propto \nu ^{-\alpha }$ (see, e.g., Pearson and Zensus 1987).
Due to the Doppler boosting effect, the flux of an approaching component
will be obviously amplified, while that of a receding component will be
significantly reduced. Therefore, one expects to observe many approaching
components rather than receding ones, and this is true. In fact, we find the
core-jet rather than twin-jet structure for the great majority of AGNs
observed to date. However, several twin-jet sources were detected recently,
both inside and outside our galaxy (those inside the Galaxy are always
called microquasars). Some were measured with proper motions of ejected
components (see, e.g., Mirabel and Rodriguez 1994, Taylor and Vermeulen
1997), while others were detected with Doppler shifted velocities (see,
e.g., Crampton et al. 1987).

It is interesting that proper motions of twin-jets can be used to estimate
the Hubble constant. This idea can be traced back to as early as 1980s (see,
Marscher and Broderick 1982). It is not until the recent discovery of proper
motions that the idea of using this method can be realized. The most
successful one was achieved by Taylor and Vermeulen (1997). The data of
proper motions can also be used to estimate the pattern speed and the angle
to the line of sight of the jet if the distance of the source is known.
However, the flow speed in the jet and the angle to the line of sight of the
jet can be well determined by Doppler shifted velocities alone, without
depending on any knowledge of the distance of the source. As a consequence,
one can combine this result together with proper motions to determine the
distance without referring to any ``standard'' quantities (see, Mirabel and
Rodriguez 1994). In this paper, we make a preview study of combining the
observations of proper motions and Doppler shifted velocities of twin-jets
to determine the distance of sources, trying to find out if this method is
realizable with current technique.

\section{The method}

According to the theory of relativity, the apparent transverse velocity of a
component is related to the angle to the line of sight and its speed (Rees
1966, 1967). For a twin-jet, the apparent transverse velocities of the
approaching and receding components follow 
\begin{equation}
(v_{app})_A\equiv \mu _AD=\frac{\beta _p\sin \theta }{1-\beta _p\cos \theta }%
c\qquad \qquad (\theta \leq \frac \pi 2),
\end{equation}
\begin{equation}
(v_{app})_R\equiv \mu _RD=\frac{\beta _p\sin \theta }{1+\beta _p\cos \theta }%
c\qquad \qquad (\theta \leq \frac \pi 2),
\end{equation}
respectively, where $\beta _p$ is the pattern speed (in the units of $c$) of
the approaching and receding components, $\theta $ is the angle to the line
of sight and $D$ is the distance of the source, while $\mu _A$ and $\mu _R$
are the proper motions of the approaching and receding components
respectively. For extragalactic sources, $D$ should be replaced by $%
(1+z)D_\theta $, where $D_\theta $ and $z$ are the angular distance and the
redshift of the source, respectively. Combining these two equations, one
gets a relation for determining the distance as 
\begin{equation}
D=\frac c{2\mu _A\mu _R}\sqrt{\beta _p^2(\mu _A+\mu _R)^2-(\mu _A-\mu _R)^2}.
\end{equation}

The Doppler shifted velocity of a component is defined as 
\begin{equation}
v_{Dop}\equiv \frac{\lambda -\lambda _0}{\lambda _0}c,
\end{equation}
where $\lambda $ is the wavelength measured by the observer and $\lambda _0$
is its proper value. Applying the Doppler effect we find 
\begin{equation}
v_{Dop}=(\frac{1-\beta _f\cos \theta }{\sqrt{1-\beta _f^2}}-1)c,
\end{equation}
where $\beta _f$ is the flow speed in the jets. Thus, the Doppler shifted
velocities of the approaching and receding components of a twin-jet should
be 
\begin{equation}
v_A=(\frac{1-\beta _f\cos \theta }{\sqrt{1-\beta _f^2}}-1)c\qquad \qquad
(\theta \leq \frac \pi 2),
\end{equation}
\begin{equation}
v_R=(\frac{1+\beta _f\cos \theta }{\sqrt{1-\beta _f^2}}-1)c\qquad \qquad
(\theta \leq \frac \pi 2),
\end{equation}
respectively. It is clear that $v_R$ must be positive, while $v_A$ can be
both positive and negative (if $\cos \theta <(1-\sqrt{1-\beta ^2})/\beta $,
then $\lambda >\lambda _0$, $v_A$ is positive; if $\cos \theta >(1-\sqrt{%
1-\beta ^2})/\beta $, then $\lambda <\lambda _0$, $v_A$ is negative). The
above two equations lead to 
\begin{equation}
\beta _f=\frac{\sqrt{(4c+v_A+v_R)(v_A+v_R)}}{2c+v_A+v_R}.
\end{equation}

From (1) and (2) one has (see also Mirabel and Rodriguez 1994) 
\begin{equation}
\beta _p\cos \theta =\frac{\mu _A-\mu _R}{\mu _A+\mu _R}\qquad \qquad
(\theta \leq \frac \pi 2)
\end{equation}
and from (6) and (7) one gets 
\begin{equation}
\beta _f\cos \theta =\frac{v_R-v_A}{2c+v_A+v_R}\qquad \qquad (\theta \leq
\frac \pi 2).
\end{equation}
Assuming 
\begin{equation}
\beta _p=\beta _f,
\end{equation}
then we have 
\begin{equation}
\frac{\mu _A-\mu _R}{\mu _A+\mu _R}=\frac{v_R-v_A}{2c+v_A+v_R}.
\end{equation}
Combining (3) and (8) and applying (11) and (12), one finds that 
\begin{equation}
D=\frac{c(\mu _A-\mu _R)}{\mu _A\mu _R}\frac{\sqrt{v_Av_R+cv_A+cv_R}}{v_R-v_A%
}.
\end{equation}
It shows, if the proper motions $\mu _A$ and $\mu _R$ and Doppler shifted
velocities $v_A$ and $v_R$ of the approaching and receding components of a
twin-jet are measured, then the distance of the source will be well
determined. Due to its simpleness, one might desire to observe proper
motions and Doppler shifted velocities of twin-jet sources and then adopt
the above equation to determined the distance of the source.

As pointed out by Mirabel and Rodriguez (1994), when $\beta _p=\beta
_f=\beta $, there will be only three unknown quantities $D$, $\beta $ and $%
\theta $, then it is only necessary to measure three out of the four
possible observables $\mu _A$, $\mu _R$, $v_A$ and $v_R$, which are related
by (12), suggesting that only three of them are independent. Applying (12)
to (13), one can have different forms of $D$ as various functions of any
three of the four observables.

For instance, if only $\mu _A$, $v_A$ and $v_R$ are measured, we simply
cancel $\mu _R$ from (12) and (13) and then have 
\begin{equation}
D=\frac c{\mu _A}\frac{\sqrt{v_Av_R+cv_A+cv_R}}{c+v_A},
\end{equation}
here $\mu _A$, $v_A$ and $v_R$ are independent.

In order to investigate the feasibility of this method, we consider the
effect from the uncertainties of the measured values of proper motions and
Doppler shifted velocities. Let the uncertainties of $\mu _A$, $v_A$ and $v_R
$ be $\Delta \mu _A$, $\Delta v_A$ and $\Delta v_R$, respectively. From
equation (14) we find the relative uncertainties of $D$ caused by $\Delta
\mu _A$, $\Delta v_A$ and $\Delta v_R$ are 
\begin{equation}
\frac{(\Delta D)_{\Delta \mu _A}}D=\frac{\Delta \mu _A}{\mu _A},
\end{equation}
\begin{equation}
\frac{(\Delta D)_{\Delta v_A}}D=\frac{\left|
[c^2-(v_Av_R+cv_A+cv_R)]v_A\right| }{2(c+v_A)(v_Av_R+cv_A+cv_R)}\frac{\Delta
v_A}{v_A},
\end{equation}
and 
\begin{equation}
\frac{(\Delta D)_{\Delta v_R}}D=\frac{v_Av_R+cv_R}{2(v_Av_R+cv_A+cv_R)}\frac{%
\Delta v_R}{v_R},
\end{equation}
respectively. The total uncertainty of $D$ is simply expressed by 
\begin{equation}
\Delta D=\sqrt{(\Delta D)_{\Delta \mu _A}^2+(\Delta D)_{\Delta
v_A}^2+(\Delta D)_{\Delta v_R}^2}.
\end{equation}

Equation (15) shows that, in any cases, the uncertainty of $D$ caused by $%
\Delta \mu _A$ is in the same order of the latter. Taking $\beta _f=\beta $
and applying (6) and (7) one finds that 
\begin{equation}
\frac{(\Delta D)_{\Delta v_A}}D=\frac{\left| [\beta ^2\sin ^2\theta
-(1-\beta ^2)](1-\beta \cos \theta -\sqrt{1-\beta ^2})\right| }{2\beta
^2\sin ^2\theta (1-\beta \cos \theta )}\frac{\Delta v_A}{v_A}
\end{equation}
and 
\begin{equation}
\frac{(\Delta D)_{\Delta v_R}}D=\frac{(1-\beta \cos \theta )(1+\beta \cos
\theta -\sqrt{1-\beta ^2})}{2\beta ^2\sin ^2\theta }\frac{\Delta v_R}{v_R}.
\end{equation}
When $\beta \simeq 1$, then $(\Delta D)_{\Delta v_A}/D\simeq (1/2)\Delta
v_A/v_A$ and $(\Delta D)_{\Delta v_R}/D\simeq (1/2)\Delta v_R/v_R$, showing
that the uncertainty of $D$ caused by $\Delta v_A$ and $\Delta v_R$ are in
the same order of them. If all the uncertainties of $\mu _A$, $v_A$ and $v_R$
are within $10\%$, then the uncertainty of $D$ would be well within $13\%$.

\section{Discussion}

In last section, we assume that the flow speed equals the pattern speed and
the angles to the line of sight used in the proper motion and Doppler
shifted velocity equations are the same. However, some of these assumptions
may not be true. For example, from VLBI proper motion studies of
extragalactic sources we know that the flow speed and the pattern speed are
not always the same.

Assuming the jet and counter-jet are in opposite directions, one can
determine $\beta _f$ by applying equation (8) and determine $\theta $ by 
\begin{equation}
\cos \theta =\frac{v_R-v_A}{\sqrt{(4c+v_A+v_R)(v_A+v_R)}}\qquad \qquad
(\theta \leq \frac \pi 2).
\end{equation}
Meanwhile, $\beta _p$ is determined by 
\begin{equation}
\beta _p=\frac{\mu _A-\mu _R}{\mu _A+\mu _R}\frac{\sqrt{(4c+v_A+v_R)(v_A+v_R)%
}}{v_R-v_A}
\end{equation}
and $D$ can be determined by applying equation (13). [Combining (3) and (22)
will lead to (13), suggesting that, even $\beta _p\neq \beta _f$, (13) is
still valid.] It shows clearly that with the four observables $\mu _A$, $\mu
_R$, $v_A$ and $v_R$ one can uniquely determine the four unknown quantities $%
D$, $\beta _p$, $\beta _f$, and $\theta $. If only three out of the four
possible observables are measured and one assumes $\beta _p=\beta _f=\beta $%
, then $D$ can also be determined, as suggested by Mirabel and Rodriguez
(1994). But if so, one might miss to observe the possible difference between
the two speeds. Thus, we suggest that, if possible, all the four observables
should be measured.

Many sources (both galactic and extragalactic) exhibit helical jet structure
which means that the angles to the line of sight for the jet and its
counter-jet are not necessarily the same. Let the angle to the line of sight
for the receding component be $\pi -\theta ^{\prime }$, with 
\begin{equation}
\theta ^{\prime }=\theta +\Delta \theta ,
\end{equation}
where, $\theta $ is the angle to the line of sight for the approaching
component and $\Delta \theta $ is the deviation of the angle for the
receding one (which can be positive or negative). In this situation, $\theta 
$ in equation (2) should be replaced by $\theta ^{\prime }$ and the
replacement would yield 
\begin{equation}
D=\frac c{\mu _R}\frac{\beta \sin (\theta +\Delta \theta )}{1+\beta \cos
(\theta +\Delta \theta )}\qquad \qquad (\theta \leq \frac \pi 2).
\end{equation}
Taking $\Delta \theta $ as a small value, we extend this equation to the
first order of $\Delta \theta $ and then get 
\begin{equation}
D=\frac c{\mu _R}\frac{\beta \sin \theta }{1+\beta \cos \theta }[1+\frac{%
\beta +\cos \theta }{(1+\beta \cos \theta )\sin \theta }\Delta \theta 
]\qquad \qquad (\theta \leq \frac \pi 2).
\end{equation}
For a twin-jet close to the plane of the sky, $\theta \simeq \pi /2$, then $%
(\beta +\cos \theta )/[(1+\beta \cos \theta )\sin \theta ]\simeq \beta $. It
shows that if $\left| \Delta \theta \right| <0.1$ ($0.1$ corresponds to $%
5.7^{\circ }$), the deviation of the distance caused by $\Delta \theta $
will be less than $10\%$ (note that $\beta <1$). So, for the requirement of $%
10\%$ uncertainty of the distance, the deviation of the angle to the line of
sight of the receding component of a twin-jet close to the plane of the sky
is allowed to be $5.7^{\circ }$.

\section{Potential observational targets}

For the convenience of choosing sources to observe, we list some sources as
potential targets.

(a) Sources with their proper motions available

As for recent observational targets, we preferentially suggest the following
sources since their proper motions have been detected and their jets seem to
be quite close to the plane of the sky.

(1) Sources in the Galaxy

GRS 1915+105 (Mirabel and Rodriguez 1994, Fender et al. 1999): $\mu
_A=23.6\pm 0.5$ mas day$^{-1}$, $\mu _R=10.0\pm 0.5$ mas day$^{-1}$

GRO J1655-40 (Hjellming and Rupen 1995): $\mu _A=54$ mas day$^{-1}$, $\mu
_R=45$ mas day$^{-1}$

XTE J1819-284 (Hjellming et al. 1999): $\mu _A=500$ mas day$^{-1}$, $\mu
_R=200$ mas day$^{-1}$

(2) Extragalactic sources

1146+596 (NGC 3894) (Taylor et al. 1998): $z=0.01085$, $\mu _A=0.26\pm 0.05$
mas yr$^{-1}$, $\mu _R=0.19\pm 0.05$ mas yr$^{-1}$

1946+708 (Taylor and Vermeulen 1997): $z=0.101$, $\mu _A=0.117\pm 0.020$ mas
yr$^{-1}$, $\mu _R=0.053\pm 0.020$ mas yr$^{-1}$

(b) Sources with their Doppler shifted velocities available

The following sources are all inside the Galaxy. They might be chosen to
observe proper motions since their Doppler shifted velocities are available.

RX J0019.8+2156 (Cowley et al. 1998): $v_R=712\pm 35$ km s$^{-1}$, $%
v_A=-690\pm 35$ km s$^{-1}$

RX J0513-69 (Southwell et al. 1996; Cowley et al. 1998): $v_R=4000$ km s$%
^{-1}$, $v_A=-3700$ km s$^{-1}$

CAL 83 (Crampton et al. 1987; Cowley et al. 1998): $v_R\sim 2450$ km s$^{-1}$%
, $v_A\sim -690$ km s$^{-1}$

(c) Potential sources

Listed in the following are some potential sources probably to be
interesting targets. Some of them are convinced to be twin-jet sources while
others might probably possess the twin-jet structure.

(1) Potential sources in the Galaxy

XTE J0421+560 (CI Cam radio source) (Hjellming and Mioduszewski 1998;
Hjellming et al. 2000): $\mu _A=\mu _R=26$ mas day$^{-1}$

SAX J1819.3-2525 (In't Zand et al. 2000): $\mu _A\simeq \mu _R\in [224,806]$
mas day$^{-1}$

XTE J1748-288 (Rupen and Hjellming 1998; Hjellming et al. 2000): $\mu _A+$ $%
\mu _R\geq [20,40]$ mas day$^{-1}$

4U 1630-47 (Hjellming et al. 2000)

XTE J1550-564 (Hjellming et al. 2000)

XTE J1739-278 (Hjellming et al. 2000)

XTE J1806-246 (Hjellming et al. 2000)

XTE J1819-245 (V4641, Sgr) (Hjellming et al. 2000)

XTE J1859+226 (Hjellming et al. 2000)

XTE J2012+381 (Hjellming et al. 2000)

(2) Potential extragalactic sources

0316+413 (NGC 1275, 3C 84) (Marr et al. 1989, Vermeulen et al. 1994): $%
z=0.0172$, $\mu _A=0.58$ mas yr$^{-1}$

NGC 1052 (Kellermann et al. 1999): $z=0.0049$, $\mu _A+$ $\mu _R=1.3$ mas yr$%
^{-1}$

NGC 4258 (Burbidge 1995; Uzernoy 1996): $z_1=0.398$, $z_2=0.653$

3C 338 (Giovannini et al. 1998): $z=0.03023$, $\mu _A\simeq $ $\mu _R\simeq
0.30-0.33$ mas yr$^{-1}$

Centaurus A (Tingay et al. 1998; Dhawan et al. 1998): $\mu _A=0.4$ mas yr$%
^{-1}$

\section{Conclusions}

In this paper, we discuss the possibility of using observational data of
twin-jets to determine the distance of sources inside and outside our
galaxy. It is known that Doppler shifted velocities of a twin-jet can solely
determine the flow speed of components of the jet and the angle to the line
of sight, while proper motions of the components can be used to determine
the distance of the source when the pattern speed of the components or the
angle to the line of sight is known. Therefore, one can observe proper
motions as well as Doppler shifted velocities of a twin-jet and then
determine the distance of the source without referring to any standard
quantities such as standard candles or rods. We investigate the feasibility
of this method by considering the uncertainty of the distance caused by the
uncertainties of the measured values of proper motions and Doppler shifted
velocities. It shows that, when the motion of the components of the jet is
relativistic, then the uncertainty of the distance is within the same order
of the uncertainties of the measured values of proper motions and Doppler
shifted velocities. For example, when assuming the pattern speed equals the
flow speed, for $10\%$ uncertainties of the measured quantities, the
uncertainty of the distance caused by them would be well within $13\%$.

As shown in last section, the uncertainty of proper motions measured inside
the Galaxy can be as small as $2\%$ (see the source of GRS 1915+105); that
measured outside the Galaxy can be as small as $17\%$ (see the source of
1946+708); while the uncertainty of Doppler shifted velocities measured
inside the Galaxy can be as small as $5\%$ now (see the source of RX
J0019.8+2156). It is obvious that, with current technique, it is possible
now to observe proper motions and Doppler shifted velocities of twin-jets
and then to determine the distance of the sources within a quite satisfied
order of uncertainty.

The advantage of this method is of course that it does not refer to any
standard quantities such as standard candles or rods. The disadvantage is
that there have been only a few twin-jet sources detected, and it is hard to
measure both the proper motion and the Doppler shifted velocity of a source
in the same time since the two quantities are currently observed in
different wavelengths. To overcome the later disadvantage, the method used
in observing NGC 4258 (Herrnstein et al. 1999) might be desired to be
applied to the known twin-jet sources. If so, we will be able to determine
the distance of a source with the well-defined method.

For the convenience of choosing sources to observe, some sources as
potential targets are also listed in this paper. For these sources, some are
known to be twin-jet ones while others might possess the twin-jet structure.

\vspace{20mm}

\begin{center}
{\bf {\Large ACKNOWLEDGMENTS}}\\
\end{center}

We thank Max-Planck-Institut fur Radioastronomie as parts of this work were
done by Dr. Yi-Ping Qin when he was a scientific guest there. Thanks are
also given to Professor G. Weigelt for his helpful advice. This work was
supported by the United Laboratory of Optical Astronomy and Laboratory of
Cosmic Ray and High Energy Astrophysics, CAS, and the Natural Science
Foundation of Yunnan.

\newpage

\begin{center}
{\bf {\Large REFERENCES}}\\
\end{center}

\begin{verse}
Brach, D. 1992, ApJ, 392, 35\\

Burbidge, E. M. 1995, A\&A, 298, L1\\

Cowley, A. P. et al. 1998, ApJ, 504, 854\\

Crampton, D. et al. 1987, ApJ, 321, 745\\

Dhawan, V., Kellermann, K., and Romney, J. D. 1998, ApJ, 498, L111\\

Fender, R. P., Garrington, S. T., McKay, D. J. et al. 1999, MNRAS, 304, 865\\

Giovannini, G. et al. 1998, ApJ, 493, 632\\

Herrnstein, J. R. et al. 1999, Nature, 400, 539\\

Hjellming, R. M., and Mioduszewski, A. J. 1998, IAU Circular No. 6872\\

Hjellming, R. M., and Rupen, M. P. 1995, Nature, 375, 464\\

Hjellming, R. M., Rupen, M. P., Mioduszewski, A. J. 2000, Rossi2000:
Astrophysics with the Rossi X-ray Timing Explorer. March 22-24, 2000 at
NASA's Golddard Space Flight Center, Greenbelt, MD USA, p. E88\\

Hjellming, R. M., et al. 1999, AAS, 195, \#134.03\\

In't Zand, J. J. M. et al. 2000, A\&A, 357, 520\\

Kellermann, K. I. et al. 1999, AAS, 194, \#20.02\\

Marr, J. M., Backer, D. C., and Wright, M. C. H. et al. 1989, ApJ, 337, 671\\

Marscher, A. P., and Broderick, J. J. 1982, In Extragalactic Radio Sources,
D. S. Heeschen and C. M. Wade, eds. (Dordrecht: Reidel), p. 359\\

Mirabel, I. F., and Rodriguez, L. F. 1994, Nature., 371, 46\\

Pearson, T. J., and Zensus, J. A. 1987, in Superluminal Radio Sources, eds.
J. A. Zensus and T. J. Pearson (Cambridge: Cambridge University Press), p. 1%
\\

Rees, M. J. 1966, Nature, 211, 468\\

Rees, M. J. 1967, MNRAS, 289, 945\\

Rupen, M. P., and Hjellming, R. M. 1998, IAU Circular No. 6938\\

Southwell, K. A. et al. 1996, ApJ, 470, 1065\\

Taylor, G. B., and Vermeulen, R. C. 1997, ApJ, 485, L9\\

Taylor, G. B., Worbel, J. M., and Vermeulen, R. C. 1998, ApJ, 498, 619\\

Tingay, S. J. et al. 1998, ApJ, 115, 960\\

Urry, C. M., and Padovani, P. 1995, PASP, 107, 803\\

Uzernoy, L. M. 1996, AAS, 188, \#16.04\\

Vermeulen, R. C., Readhead, A. C. S., and Backer, D. C. 1994, ApJ, 430, L41\\
\end{verse}

\end{document}